\documentstyle{mn}

\newif\ifAMStwofonts

\ifoldfss
  \ifCUPmtlplainloaded \else
    \NewTextAlphabet{textbfit} {cmbxti10} {}
    \NewTextAlphabet{textbfss} {cmssbx10} {}
    \NewMathAlphabet{mathbfit} {cmbxti10} {} 
    \NewMathAlphabet{mathbfss} {cmssbx10} {} 
  \fi
  \ifAMStwofonts
    \ifCUPmtlplainloaded \else
      \NewSymbolFont{upmath} {eurm10}
      \NewSymbolFont{AMSa} {msam10}
      \NewMathSymbol{\upi}     {0}{upmath}{19}
      \NewMathSymbol{\umu}     {0}{upmath}{16}
      \NewMathSymbol{\upartial}{0}{upmath}{40}
      \NewMathSymbol{\leqslant}{3}{AMSa}{36}
      \NewMathSymbol{\geqslant}{3}{AMSa}{3E}

      \let\leq=\leqslant \let\le=\leqslant
      \let\geq=\geqslant \let\ge=\geqslant
    \fi
  \fi
\fi 

\ifnfssone
  \newmathalphabet{\mathit}
  \addtoversion{normal}{\mathit}{cmr}{m}{it}
  \addtoversion{bold}{\mathit}{cmr}{bx}{it}
  \newmathalphabet{\mathbfit} 
  \addtoversion{normal}{\mathbfit}{cmr}{bx}{it}
  \addtoversion{bold}{\mathbfit}{cmr}{bx}{it}
  \newmathalphabet{\mathbfss} 
  \addtoversion{normal}{\mathbfss}{cmss}{bx}{n}
  \addtoversion{bold}{\mathbfss}{cmss}{bx}{n}
  \ifAMStwofonts
    \ifCUPmtlplainloaded \else
      \UseAMStwoboldmath
      \makeatletter
      \new@mathgroup\upmath@group
      \define@mathgroup\mv@normal\upmath@group{eur}{m}{n}
      \define@mathgroup\mv@bold\upmath@group{eur}{b}{n}
      \edef\UPM{\hexnumber\upmath@group}
      \new@mathgroup\amsa@group
      \define@mathgroup\mv@normal\amsa@group{msa}{m}{n}
      \define@mathgroup\mv@bold\amsa@group{msa}{m}{n}
      \edef\AMSa{\hexnumber\amsa@group}
      \makeatother
      \mathchardef\upi="0\UPM19
      \mathchardef\umu="0\UPM16
      \mathchardef\upartial="0\UPM40
      \mathchardef\leqslant="3\AMSa36
      \mathchardef\geqslant="3\AMSa3E

      \let\leq=\leqslant \let\le=\leqslant
      \let\geq=\geqslant \let\ge=\geqslant
    \fi
  \fi
\fi 

\ifnfsstwo
  \DeclareMathAlphabet{\mathbfit}{OT1}{cmr}{bx}{it}
  \SetMathAlphabet\mathbfit{bold}{OT1}{cmr}{bx}{it}
  \DeclareMathAlphabet{\mathbfss}{OT1}{cmss}{bx}{n}
  \SetMathAlphabet\mathbfss{bold}{OT1}{cmss}{bx}{n}
  \ifAMStwofonts
    \ifCUPmtlplainloaded \else
      \DeclareSymbolFont{UPM}{U}{eur}{m}{n}
      \SetSymbolFont{UPM}{bold}{U}{eur}{b}{n}
      \DeclareSymbolFont{AMSa}{U}{msa}{m}{n}
      \DeclareMathSymbol{\upi}{0}{UPM}{"19}
      \DeclareMathSymbol{\umu}{0}{UPM}{"16}
      \DeclareMathSymbol{\upartial}{0}{UPM}{"40}
      \DeclareMathSymbol{\leqslant}{3}{AMSa}{"36}
      \DeclareMathSymbol{\geqslant}{3}{AMSa}{"3E}

      \let\leq=\leqslant \let\le=\leqslant
      \let\geq=\geqslant \let\ge=\geqslant
    \fi
  \fi
\fi 

\ifCUPmtlplainloaded \else
  \ifAMStwofonts \else 
    \def\upi{\pi}
    \def\umu{\mu}
    \def\upartial{\partial}
  \fi
\fi

\title{Protostellar Collapse Induced by Compression}
\author[P. Hennebelle, A. P. Whitworth, P. P. Gladwin \& Ph. Andr\'e]
       {P. Hennebelle,$^1$\thanks{Patrick.Hennebelle@astro.cf.ac.uk} 
        A. P. Whitworth,$^1$\thanks{ant@astro.cf.ac.uk}
        P. P. Gladwin,$^1$
        Ph. Andr\'e,$^2$\thanks{pandre@discovery.saclay.cea.fr}\\
        $^1$Department of Physics \& Astronomy, Cardiff University, 
        PO Box 913, 5 The Parade, Cardiff CF24 3YB, Wales, UK\\
        $^2$CEA, DSM, DAPNIA, Service d'Astrophysique, C. E. Saclay, 
        F-91191 Gif-sur-Yvette Cedex, France}
\date{Accepted.
      Received;
      in original form}

\pagerange{\pageref{firstpage}--\pageref{lastpage}}
\pubyear{2001}

\newcommand{\xt}{_{\rm ext}}
\newcommand{\nb}{_{\rm neib}}
\newcommand{\bg}{_{\rm o}}

\begin{document}

\maketitle

\label{firstpage}

\begin{abstract}

We present numerical simulations of the evolution of low-mass, 
isothermal, molecular cores which are subjected to an increase in 
external pressure $P\xt$. If $P\xt$ increases very slowly, the core approaches 
instability quite quasistatically. However, for larger (but still 
quite modest) $dP\xt/dt$ a compression wave is driven into the 
core, thereby triggering collapse from the outside in. If collapse 
of a core is induced by increasing $P\xt$, this has a 
number of interesting consequences. (i) The density profile is 
approximately flat in the centre during the prestellar phase (i.e. 
before the compression wave converges on the centre creating the central 
protostar). (ii) During the prestellar phase there 
are (subsonic) inward velocities in the outer layers of the 
core, whilst the inner parts are still approximately at rest. (iii) There 
is an initial short phase of rapid accretion (notionally the Class 0 phase), 
followed by a longer phase of slower accretion (the Class I 
phase). All these features accord well with observation, 
but are at variance with the predictions of the standard theory of 
star formation based on the inside-out collapse of a singular isothermal 
sphere. We note that the setting up of a coherent inward velocity field 
appears to be a generic feature of compression waves; and we speculate 
that interactions and interference between such velocity fields may play 
a crucial r\^ole in initiating the fragmentation of cores and the genesis 
of multiple star systems.

\end{abstract}


\section{Introduction}

The initial conditions for star formation are in general rather 
weakly constrained. The observations (e.g. Andr\'e, Ward-Thompson, 
\& Barsony 2000; Myers, Evans, \& Ohashi 2000) are limited by telescope 
resolution and confusion, and their interpretation is hindered 
by the complexities and uncertainties of gas-phase abundances, 
excitation conditions, and radiation transport. However, a picture is 
emerging which suggests that star formation is sometimes triggered 
rather impulsively.

Molecular-line mapping by Myers and collaborators 
(e.g. Myers \& Benson 1983; Benson \& Myers 1989) and subsequent 
searches by Clemens \& Barvainis (1988), Bourke, Hyland \& 
Robinson (1995), Jessop \& Ward-Thompson (2000), have established 
a large sample of dense cores which appear to be the sites of 
ongoing or imminent star formation. Many of these cores contain 
IRAS sources (Beichman et al. 1986), and are therefore presumed 
to have already formed protostars. Those which do not contain 
IRAS sources are termed starless cores.

Submillimeter mapping of these starless cores by Ward-Thompson 
et al. (1994, 1999) has identified a subset which, on the basis of 
their virial ratios and relatively high central densities, 
are likely to be in a state of imminent 
or on-going contraction. These starless cores are therefore termed 
prestellar (strictly pre-protostellar). Submillimeter mapping of 
prestellar cores (Ward-Thompson et al. 1994, Andr\'e et al. 1996, 
Ward-Thompson et al. 1999) suggests that, if the emitting dust is isothermal, 
the cores have rather flat density profiles in the centre. Specifically, 
$\eta \equiv - d \ell n [\rho] / d \ell n [r]$ is in the range 0 to 1,  
in the innermost few thousand AU. 
This conclusion may be somewhat weaker if, as seems 
likely (Jessop \& Ward-Thompson 2001; Evans et al. 2001; Zucconi, 
Walmsley \& Galli 2001; Ward-Thompson, Andr\'e \& Kirk, 2001), the 
dust temperature decreases towards the centre of a core. However, 
mid-infrared absorption measurements of prestellar cores 
-- which are insensitive to the temperature profile -- also 
indicate flat inner density profiles (Bacmann et al. 2000). 
Additionally, mid-infrared observations suggest that beyond 
$\sim$ 10,000 AU the density profile may steepen to 
$\eta \ga 4\,$ (Abergel et al. 1996, Bacmann et al. 2000).
This steepening of the outer envelope appears to occur at smaller 
radii in relatively close-packed protoclusters like $\rho$ Ophiuchi 
than in regions of distributed star formation such as Taurus; viz. at 
$\sim$ 3,000 AU in $\rho$ Ophiuchi, and at $\sim$ 15,000 AU in Taurus 
(Motte \& Andr\'e 2001).

The asymmetric self-absorbed 
molecular-line profiles observed in some prestellar cores (Tafalla 
et al. 1998, Williams et al. 1999, Lee, Myers \& Tafalla 1999, 
Gregersen \& Evans 2000) suggest that they are indeed 
already collapsing. In particular, the detailed analyses of 
L1544 by Tafalla et al. (1998) and Williams et al. (1999) imply 
that the inner parts of the core are relatively stationary, and an 
approximately uniform velocity field has been established in the 
outer layers. This is very reminiscent of the velocity fields which 
are set up by inward-propagating compression waves in similarity 
solutions for contracting isothermal spheres (Whitworth \& Summers 
1985). It is this apparent similarity which we explore in the present 
paper.

The prestellar phase terminates as soon as a star-like object 
forms at the centre of the dense core, and the core then becomes a 
Class 0 protostar. Conceptually, the Class 0 protostellar phase 
terminates once the extended envelope 
contains less than half the total mass of the original core, 
i.e. more than half the mass is in the central star-like 
object(s) plus attendant disc(s) (Andr\'e, Ward-Thompson 
\& Barsony 1993, 2000); the Class I phase then begins 
(Lada \& Wilking 1984, Lada 1987, Andr\'e \& Montmerle 1994).
The Class I protostellar phase terminates when most of the 
envelope has been accreted or dissipated, revealing a classical 
T Tauri star (CTTS) accreting from a residual circumstellar disc, i.e. 
a Class II object. Once the inner disc has been dissipated, 
the accretion rate is greatly reduced and the source becomes 
a weak-lined T Tauri star (WTTS) or Class III object.

On the basis of statistical arguments (i.e. source numbers and 
a presumed constant star formation rate) it is inferred (Beichmann 
et al. 1986, 
Andr\'e, Ward-Thompson \& Barsony 2000) 
that the prestellar phase lasts, typically, $10^6$ to $10^7$ 
years. In close-packed protoclusters like $\rho$ Ophiuchi, the 
Class 0 phase appears to last a few times $10^4$ years, 
and is characterized by powerful collimated outflows indicative of 
rapid accretion, $\ga 10^{-5} \, M_\odot \, \mbox{year}^{-1}$ 
(Bontemps et al. 1996). In distributed star formation regions 
like Taurus, the duration of the Class~0 phase appears to be longer,  
$\sim 10^5$ years, and the accretion rates lower, $\ga 2 \times 
10^{-6} \, M_\odot \, \mbox{year}^{-1}$ (Motte \& Andr\'e 2001). 
The Class I phase appears to last $\sim 2 \times 10^5$ years
(e.g. Greene et al. 1994, Kenyon \& Hartmann 1995) and is 
characterized by 
slower accretion, $\la 10^{-6} \, M_\odot \, \mbox{year}^{-1}$, 
and weaker less collimated outflows (Bontemps et al. 1996).
Together, the Class II and 
Class III phases appear to last $\ga 10^7$ years, 
but seemingly the transition from Class II (CTTS) to Class III (WTTS) 
is very short and can occur at any time; 
there are both very young WTTSs (close to the birthline, on the right of
the Hertzsprung-Russell Diagram) and very old CTTSs 
(approaching the Main Sequence on the left of the Hertzsprung-Russell 
Diagram) (e.g. Stahler \& Walter 1993).

These observationally inferred features of prestellar and 
protostellar evolution combine to form a reasonably coherent 
picture. However, especially in star-forming clusters, this 
picture is difficult to reconcile with the standard theory of Shu, 
Adams \& Lizano (1987) based on the inside-out collapse of a 
singular isothermal sphere. The strength of the standard theory 
is that it makes specific quantitative predictions, but some of 
these predictions are difficult to reconcile with 
observation. (i) In the standard model, prestellar cores 
should be strongly centrally condensed, $\eta \equiv - 
d \ell n [\rho] / d \ell n [r] \sim 2$ and young protostars 
should be somewhat less centrally condensed, $\eta \sim 
3/2\,$. By contrast, observations suggest that prestellar cores have 
rather flat central density profiles, $\eta \la 1$, and Class 0 
protostars have steeper ones. (ii) The standard theory predicts 
that prestellar cores are static, and that inward motions only 
develop during the protostellar phase and are initially confined 
to the central regions. In contrast, the observations imply that 
inward motions already exist during the prestellar phase, and that 
initially they are more rapid in the outer regions. (iii) The standard 
theory predicts that the accretion rate is roughly constant, 
and hence the Class 0 and Class I lifetimes should be comparable. 
Observations detect many more Class I sources than Class 0 ones -- 
although this statistical result may be compromised by the difficulty 
of measuring precisely when an object has accreted half the total mass of 
its initial core.  
(iv) Additionally, the standard theory assumes initial 
conditions which are unlikely to arise in nature, because they are 
both singular and unstable. (v) Also the standard theory has a strong 
inbuilt pre-disposition to the formation of single stars -- in stark 
contrast with the high proportion of binaries and higher multiples 
observed in young star-formation regions. Therefore, although the 
standard theory may provide a good, zeroth-order description of 
protostellar collapse in sparse, quiescent  star-formation regions 
like Taurus (cf. Motte \& Andr\'e 2001), it appears that more 
dynamical models are required to understand close-packed regions 
like $\rho$ Ophiuchi.

Foster \& Chevalier (1993) have explored how the collapse of an 
isothermal core develops if one abandons the assumption of singularity. 
Their simulations start from non-singular isothermal equilibria, i.e.
Bonnor-Ebert spheres (e.g. Bonnor 1956),  and 
collapse is then triggered by discontinuously increasing the density. 
The ensuing collapse results in supersonic inflow velocities in the 
centre of the core at the end of the prestellar phase, and in a 
marked decline of the accretion rate from the Class 0 to the Class I phase 
(see Henriksen, Andr\'e, Bontemps 1997 and Whitworth \& Ward-Thompson 2001 
for simpler, pressure-free descriptions of this evolution).

In this paper 
we pursue the consequences of non-singularity further. Our simulations 
also start from non-singular isothermal equilibrium cores, but collapse 
is then triggered by a steady increase in the external pressure (cf. Myers \& 
Lazarian 1998). Most of the observational constraints detailed above 
can be reproduced rather well by this model.

In Section 2 we describe the numerical 
method we use, and the initial and boundary conditions. In Section 
3 we present the results, and in Section 4 we discuss them. 
Section 5 summarizes our main conclusions.

\section{Numerical Method}

The simulations are performed using a Smoothed Particle 
Hydrodynamics code -- essentially that described in Turner et al. 
(1995). We use the B2-spline smoothing kernel of Monaghan \& 
Lattanzio (1985),

\begin{equation}
W(s) \; = \; \frac{1}{4 \pi} \left\{ \begin{array}{ll}
4 - 6 s^2 + 3 s^3 \,, & 0 \leq s \leq 1 \,; \\
(2-s)^3 \,, & 1 < s \leq 2 \,; \\
0 \,, & s > 2 \, . \\
\end{array} \right.
\end{equation}

\noindent and the smoothing lengths of the individual particles are adjusted 
so that the kernel encompasses ${\cal N}\nb \simeq 50 \pm 5$ 
neighbours, i.e. for particle $i$ the smoothing length $h_i$ is 
adjusted so that there are ${\cal N}\nb$ other particles $j$ 
having

\begin{equation}
\label{NEIB}
|{\bf r}_j - {\bf r}_i| \; < \; 2 \, \bar{h}_{ij} \; \equiv \; 
h_i + h_j \; .
\end{equation}

If we define $\Delta {\bf v}_{ij} \equiv {\bf v}_i - {\bf v}_j\,$; 
$\Delta {\bf r}_{ij} \equiv {\bf r}_i - {\bf r}_j\,$; 
$\bar{a}_{ij} \equiv 0.5(a_i+a_j)\,$; 
$\bar{\rho}_{ij} \equiv 0.5(\rho_i+\rho_j)\,$; 
 
\begin{eqnarray} \nonumber
\Pi_{ij} & = & \frac{- \alpha \mu_{ij} \bar{a}_{ij} + \beta 
\mu_{ij}^2}{\bar{\rho}_{ij}} \, , \\ \nonumber
 & & \\
\mu_{ij} & = & \left\{ \begin{array}{ll}
\bar{h}_{ij} {\bf v}_{ij} {\large\bf .} {\bf r}_{ij} \left( 
|{\bf r}_{ij}|^2 + \gamma \bar{h}_{ij}^2 \right)^{-1} \, , &
{\bf v}_{ij} {\large\bf .} {\bf r}_{ij} < 0 \, , \\
0 \,, & {\bf v}_{ij} {\large\bf .} {\bf r}_{ij} > 0 \, ; \\ \nonumber
\end{array} \right.
\end{eqnarray}

\noindent $W'(s) \equiv dW/ds\,$, and

\begin{equation}
W^*(s) \; = \; \int_{s'=0}^{s'=s} \, W(s') \, 4 \pi s'^2 \, ds' \,;
\end{equation}

\noindent then the equation of motion for particle $i$ is

\begin{eqnarray} \nonumber \label{MOTION}
\frac{d {\bf v}_i}{dt} & = & - \, \sum_j \left\{ 
\frac{m_j}{\bar{h}_{ij}^4} \left[ \frac{P_i}{\rho_i^2} + 
\frac{P_j}{\rho_j^2} + \Pi_{ij} \right] \right. \\ \nonumber
 & & \hspace{2.07cm} \left. W'\left( 
\frac{|\Delta {\bf r}_{ij}|}{\bar{h}_{ij}} \right) 
\frac{\Delta {\bf r}_{ij}}{|\Delta {\bf r}_{ij}|} \right\} \\
 & & - \, \sum_{j'} \left\{ m_{j'} W^*\left( 
\frac{|\Delta {\bf r}_{ij'}|}{\bar{h}_{ij'}} \right) 
\frac{\Delta {\bf r}_{ij'}}{|\Delta {\bf r}_{ij'}|^3} \right\} 
\,. \\ \nonumber
\end{eqnarray}

The first summation determines the hydrostatic and viscous 
accelerations. Since $W'(s) = 0\,$ for $s > 2\,$, this 
summation only extends over neighbours, i.e. particles 
satisfying Eqn. (\ref{NEIB}). $\Pi_{ij}$ is the artificial 
viscosity term and acts only between neighbours which are 
approaching one another. We use $\alpha = 1\,$, $\beta = 2\,$, 
and $\gamma = 0.01\,$.

The second summation determines the gravitational acceleration. 
Since $W^*(s) > 0\,$ for all $s > 0\,$, this summation extends over 
all particles. The gravitational attraction between particles 
whose kernels overlap is softened by $W^*(s)\,$, in accordance 
with Gauss's Gravitational Theorem; implicitly the gravitational 
softening length and the hydrostatic smoothing length are the 
same, as advocated by Bate \& Burkert (1997). We use an octal 
spatial tesselation tree (Barnes \& Hut 1986, Hernquist 1987) to 
find neighbours, and also to speed up the calculation of gravity. 
Thus in the second summation in Eqn. (\ref{MOTION}) $j'$ may 
represent the identifier of an individual particle or a cell in 
the tree. To evaluate this sum, the tree is walked from the root 
cell (representing the whole computational domain) downwards. 
Whenever a cell $j'$ is encountered which satisfies the 
non-opening condition

\begin{equation}
\label{OPENING}
L_{j'} \; < \; \theta_{\rm crit} |{\bf r}_i - {\bf R}_{j'}|\,,
\end{equation}

\noindent that cell contributes to the sum as a point mass, and 
all smaller cells and individual particles within that cell can be 
neglected. In Eqn. (\ref{OPENING}), $L_{j'}$ is the linear size of 
the cell under consideration, ${\bf R}_{j'}$ is the position of the 
cell's centre of mass, and $\theta_{\rm crit}$ is the maximum opening 
angle which a cell can subtend at particle $i$ without being opened. 
We use $\theta_{\rm crit} = 3^{-1/2} \simeq 0.577\,$.

The density at particle $i$ is given by

\begin{equation}
\rho_i \; = \; \sum_{j} \left\{ \frac{m_j}{\bar{h}_{ij}^3} 
W\left( \frac{|\Delta {\bf r}_{ij}|}{\bar{h}_{ij}} \right) 
\right\} \;.
\end{equation}

We use an isothermal equation of state 

\begin{equation}
a_i \; = \; a\bg \,, \hspace{1cm} P_i \; = \; a\bg^2 \rho_i \,,
\end{equation}

\noindent and so we do not need to solve an energy equation.

The code uses multiple-particle time-steps, 

\begin{equation}
\Delta t_n \; = \; \Delta t\bg \, 2^n \,, \hspace{1cm} 
n \; = \; 0, 1, 2, ... n_{\rm max} \,.
\end{equation}

\noindent The maximum possible time-step for each particle is given by

\begin{eqnarray} \nonumber
\Delta t_i^{\rm max} & = & \delta \; {\rm MIN} \left\{ 
\frac{1}{|\nabla . {\bf v}|_i} \,,\;
\frac{h_i}{|{\bf v}_i|} \,,\; \left( \frac{h_i}{|{\bf a}_i|} 
\right)^{1/2} \right. \, , \\
 & & \hspace{1.2cm} \left. \frac{h_i}{\left( 2.2 \, c_i + 1.2 \, 
{\tiny\rm MAX}_j \left\{ \mu_{ij} \right\} \right)} \right\} \, . \\ \nonumber
\end{eqnarray}

\noindent Then each particle $i$ is allocated the largest $\Delta t_n\,$ 
which is smaller than $\Delta t_i^{\rm max}\,$, i.e. 
$\Delta t_{n_i}\,$ with

\begin{equation}
n_i \; = \; {\rm INT} \left\{ \frac{\ell o g [\Delta t_i^{\rm max} 
/ \Delta t\bg]}{\ell o g [2]} \right\} \,,
\end{equation}

\noindent with the additional proviso that the current time-step must be an 
integer multiple of $\Delta t_{n_i}\,$.

At the start of a simulation, cores are modelled as truncated 
equilibrium isothermal spheres, contained by a hot rarefied 
external medium. These initial conditions are realized by 
distributing particles randomly in a cubic box with periodic 
boundary conditions, and settling them using SPH forces only 
(i.e. the terms in the first summation in Eqn. (\ref{MOTION})); 
this creates a uniform-density settled cube. Next a sphere is 
cut from the settled cube, and the particle positions within 
the sphere are stretched radially to reproduce the density 
profile of a truncated isothermal sphere. Isothermal spheres 
having different degrees of central condensation are created 
by picking different values of $\xi_{\rm b} = R\bg 
\left( 4 \pi G \rho_{\rm c} \right)^{1/2} a\bg^{-1}\,$, where 
$\xi_{\rm b}$ is the dimensionless radius of a truncated equilibrium 
isothermal sphere, $R\bg$ is the physical radius of the core ({\em not} 
the scale-length), and $\rho_{\rm c}$ is the central density in the core. 
Stability requires $\xi_{\rm b} < 6.45\,$ (Bonnor 1956).

The particles inside the sphere are then tagged as 
being {\it internal} particles, and henceforth they experience 
both SPH forces and mutual gravity, i.e. they are evolved using 
both the summations in Eqn. (\ref{MOTION}). The particles outside 
the sphere are tagged as {\it external} particles, and henceforth 
they are evolved using only SPH forces, i.e. only the first 
summation in Eqn. (\ref{MOTION}). The external particles are 
given whatever sound speed is needed for them to deliver the 
prescribed external pressure $P\xt(t)\,$ at the edge of the sphere, 
i.e. $a_i(t) = \left( P\xt(t) / \rho_i(t) \right)^{1/2}\,$. In 
addition, there is a layer of fixed {\it edge} particles at the 
boundary of the computational domain. These fixed edge particles 
have constant density $\rho_{\rm edge}\,$ and they are not evolved, 
but they exert SPH forces on the external particles and their 
sound speed is given by $a_{\rm edge} = \left( P\xt(t) / 
\rho_{\rm edge} \right)^{1/2}\,$.

At the outset, the sphere is relaxed with $P\xt$ held constant, so 
that it can settle to equilibrium. There is some variance 
between the true isothermal sphere characterized by the chosen value 
of $\xi_{\rm b}$ and the actual SPH-modelled configuration, due to 
the density smoothing at the core boundary, where the density gradient 
is steep. 
This leads to an effective external pressure which is a little higher 
than the specified one. The only significant consequence of this is that the
 onset of collapse is somewhat accelerated in the case of slow compression 
($\phi=10$). This is due to transient oscillations which for slow compression
have time to propagate inwards from the boundary and increase the density 
at the centre.

Stable equilibrium isothermal spheres constructed in this way 
are then subjected to increasing external pressure. For simplicity 
we prescribe

\begin{equation}
P\xt(t>0) \; = \; P\xt(0) \, + \, \dot{P}\bg t \,,
\end{equation}

\noindent with constant $\dot{P}\bg\,$. Rather than give $\dot{P}\bg$ 
in physical units, we define a dimensionless parameter

\begin{equation}
\phi \; = \; \frac{P\xt(0)/\dot{P}\bg} {R\bg / a\bg} \,,
\end{equation}

\noindent which is the ratio of the time-scale on which the external 
pressure doubles to the initial sound-crossing time ($R\bg$ 
is the initial radius of the core). Small $\dot{P}\bg$ corresponds 
to large $\phi\,$. Simulations have been performed with a variety 
of $\phi$ values between $\phi = 10$ (very slow compression) and 
$\phi = 0.1$ (very fast compression).

There is a simple spherical sink at the centre of the core, having 
radius 150 AU (i.e. $\sim$\,1\% of the initial core radius), and representing 
the central protostar (plus attendant disc). Any particle entering the sink 
is assimilated by it. The protostar is assumed to come into existence 
as soon as the sink starts to gain mass. Hence we can easily compute its 
mass and accretion rate, as functions of time.

We fix the core mass at {\bf $M\bg = 1\, M_\odot\,$}, the sound speed in the 
core at $a\bg = 0.2 \, {\rm km} \, {\rm s}^{-1}\,$ 
(corresponding to a cosmic mixture of molecular hydrogen, helium 
and heavy elements at $\sim 10 \, {\rm K}$), and the 
stability parameter of the initial core at $\xi_{\rm b} = 3.0\,$. 
The radius of the core is initially $\,\sim\,0.08\,{\rm pc}\,$.
The standard simulations presented in the next section are performed 
with ${\cal N}_{\rm int} \sim$ 50,000 internal particles, 
${\cal N}_{\rm ext} \sim$ 25,000 external particles, and 
${\cal N}_{\rm edge} \sim$ 20,000 edge particles; some simulations 
with fewer particles have been performed to demonstrate convergence.

\section{Results}

\begin{figure*}
\setlength{\unitlength}{1mm}
\begin{picture}(160,225)
\includegraphics{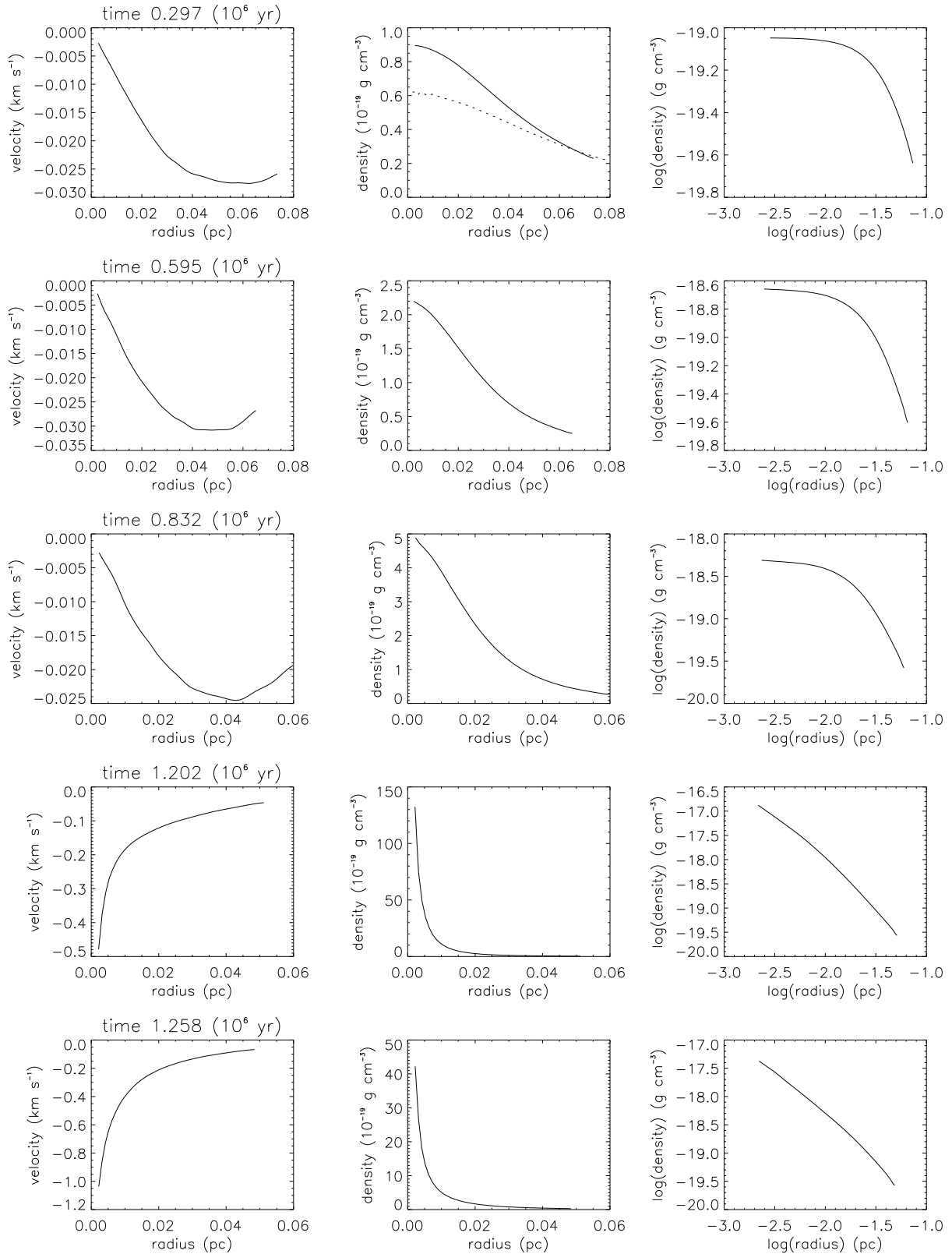}
\end{picture}
\caption{The lefthand column shows velocity profiles on a linear-linear 
scale, the central column shows density profiles on a linear-linear scale, 
and the righthand column shows density profiles on a log-log scale, for 
the case $\phi = 10.0$ (very subsonic compression), at three times during 
the prestellar phase and two in the Class 0 phase:
first row, $t = 0.30 \,\mbox{\small Myrs}$; 
second row, $t = 0.60 \,\mbox{\small Myrs}$; 
third row, $t = 0.83 \,\mbox{\small Myrs}$; 
fourth row, $t = 1.20 \,\mbox{\small Myrs}$;
fifth row, $t = 1.26 \,\mbox{\small Myrs}$.
The dashed curve in the first row represents the initial state. 
We note that with cosmic abundances, a mass-density $\rho = 10^{-19}\,
{\rm g}\,{\rm cm}^{-3}$ corresponds to a number-density 
$n_{{\rm H}_2} \simeq 2 \times 10^{-4}\,{\rm cm}^{-3}$.}
\label{phi10}
\end{figure*}

\begin{figure*}
\setlength{\unitlength}{1mm}
\begin{picture}(160,225)
\includegraphics{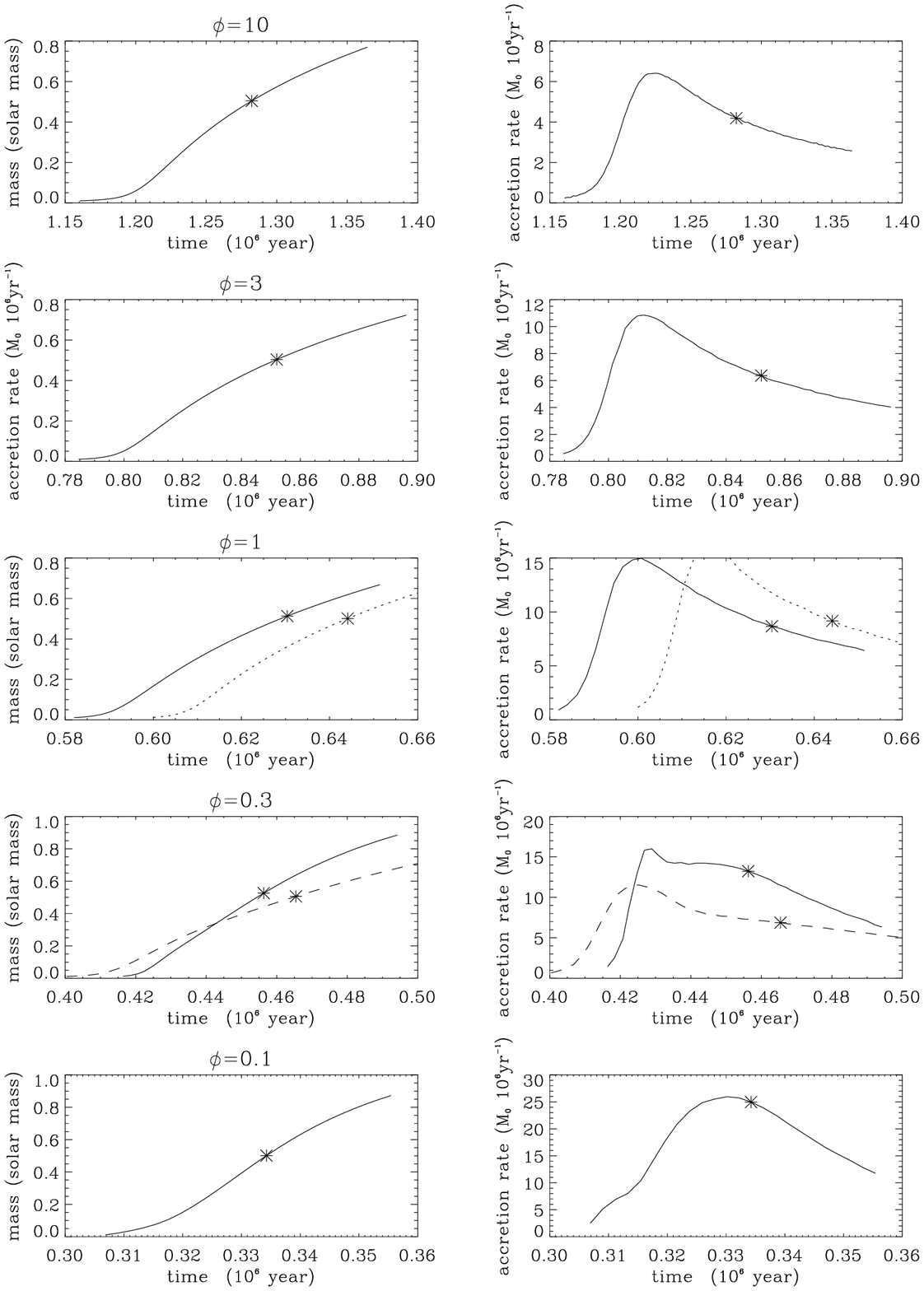}
\end{picture}
\caption{Accretion rates as a function of time. First row, $\phi=10\,$; 
second row, $\phi=3\,$; third row, $\phi=1\,$; fourth row, $\phi=0.3\,$; 
fifth row, $\phi=0.1\,$. The full curves correspond to the standard 
simulations, i.e. those in which $P\xt$ is allowed to increase indefinitely. 
The dotted curves on the third row correspond to the ``$\phi=1\,$(indefinite)'' 
case, but performed with fewer particles (i.e. ${\cal N}_{\rm int}=20,000$ 
instead of the standard ${\cal N}_{\rm int}=50,000$) in the core. The dashed 
curves on the fourth row correspond to the case $\,\phi=0.1\,$ (very fast 
compression), but where $P\xt$ is held constant once it has doubled, 
i.e.``$\phi=0.1\,$(finite)''. It is plotted with the $\phi=0.3\,$(indefinite)'' 
case on the fourth row, because the time-scale on which the compression wave 
propagates into the centre is very similar in these two cases.}
\label{accret}
\end{figure*}

Figure~\ref{phi10} shows detailed results for the $\phi = 10$ case (i.e. 
very slow compression; $P\xt$ takes ten sound-crossing times to double). 
In this case the evolution towards gravitational instability is relatively 
quasistatic, because there is time for sound waves to slop around 
re-distributing the matter. During the prestellar phase (the first three 
rows on Fig.~\ref{phi10}), the outer boundary is pushed inwards, and 
the core becomes increasingly centrally condensed; the outer parts of the 
core move slowly inwards at speeds up to $\sim 0.03 \, {\rm km} \, 
{\rm s}^{-1} (\equiv 0.15 a\bg)$, whilst matter near the centre is 
approximately stationary. Eventually the core becomes unstable and collapses; 
this is the start of the Class 0 phase. At this stage the radius of the core 
is $\sim$ 0.06 pc. As the subsequent collapse proceeds, a freefall 
velocity field ($v \propto r^{-1/2}$) develops from the inside out, steadily 
replacing the more modest approximately uniform velocity field in the outer 
layers (the fourth and fifth rows on Fig.~\ref{phi10}). The rate of 
accretion onto the central protostar (sink) is quite modest, with a maximum 
value $\sim 0.65 \times 10^{-5} \,M_\odot \, {\rm\small yr}^{-1}\,$, 
after which it decreases steadily into the Class I phase. The evolution of 
the sink mass and the accretion rate is shown on the first row of Fig.~\ref{accret}. 
The durations of the prestellar and Class 0 phases are given in Table 
1, along with an estimate of the mean cruising velocity set up by the 
passage of the compression wave.

This case is the one closest to the standard case of Foster \& Chevalier 
(1993). In their simulation collapse was initiated by taking a marginally unstable (i.e. critical) Bonnor-Ebert sphere and increasing the density 
by 10\%. As a result, contraction was driven by an imbalance between 
self-gravity and pressure. Immediately following the formation of the 
central protostar the accretion rate was very high, and subsequently it 
decreased. In the $\phi = 10$ case here, contraction is driven by compression, 
and when the resulting compression wave converges on the centre the 
accretion rate builds up to a maximum somewhat more slowly (than in 
Foster and Chevalier's simulation). In both cases the maximum accretion 
rate occurs during the Class 0 phase, and is significantly 
larger than the standard $\dot{M} \sim a\bg^3/G \sim 2 \times 10^{-6} 
M_\odot\,{\rm year}^{-1}$.

\begin{figure*}
\setlength{\unitlength}{1mm}
\begin{picture}(160,225)
\includegraphics{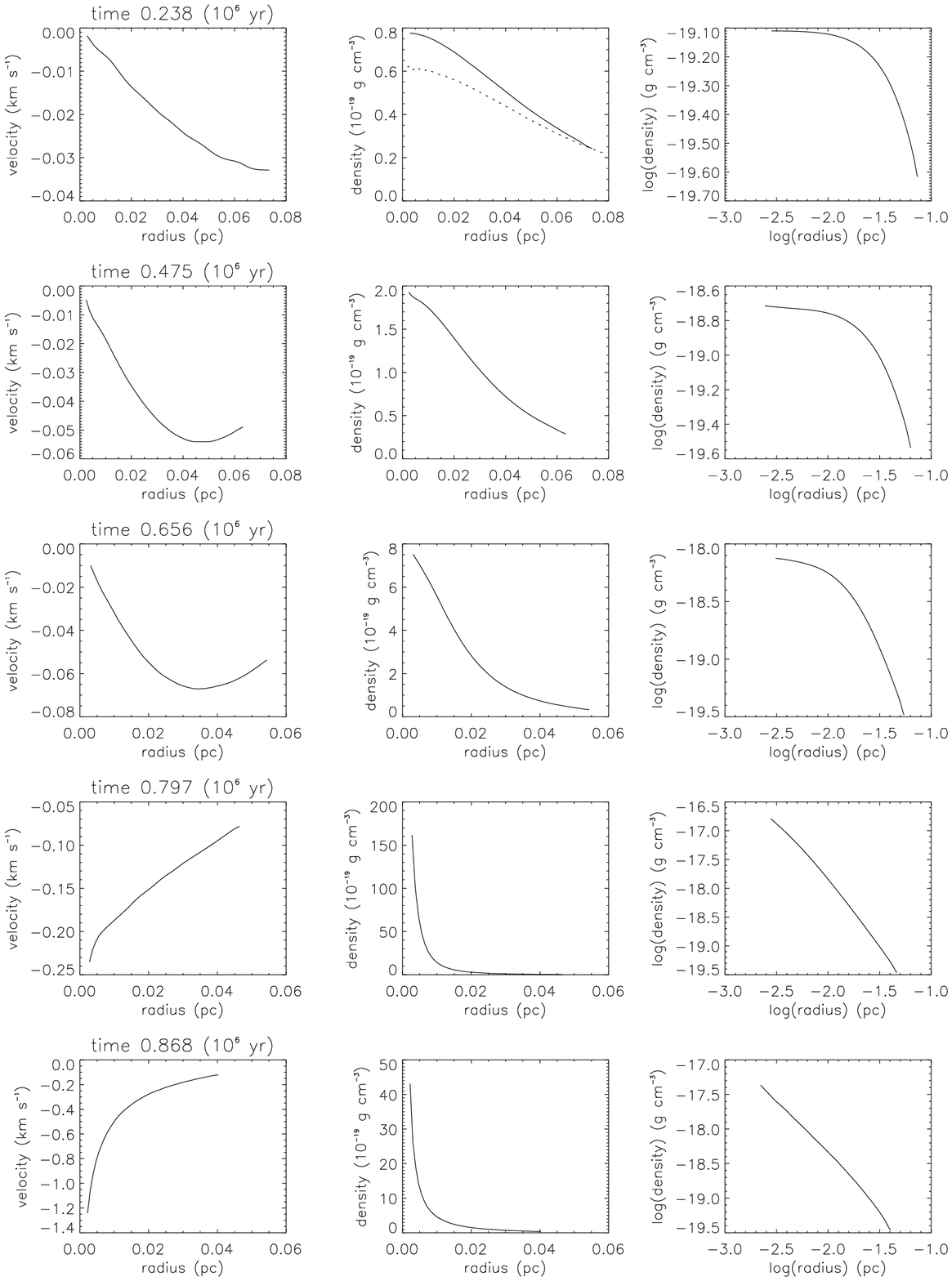}
\end{picture}
\caption{As for Fig.~\ref{phi10}, but for the case 
$\phi = 3.0$ (subsonic compression): 
first row, $t = 0.24 \,\mbox{\small Myrs}$; 
second row, $t = 0.48 \,\mbox{\small Myrs}$; 
third row, $t = 0.66 \,\mbox{\small Myrs}$; 
fourth row, $t = 0.80 \,\mbox{\small Myrs}$;
fifth row, $t = 0.87 \,\mbox{\small Myrs}$.} 
\label{phi3}
\end{figure*}

Figure~\ref{phi3} shows the case $\phi = 3$ (i.e. quite slow 
compression; $P\xt$ doubles in 3 sound-crossing times). The evolution is 
somewhat less quasistatic (than for $\phi = 10$, Fig.~\ref{phi10}), 
but basically it is very similar. The outer boundary is pushed inwards, 
and the core becomes steadily more centrally 
condensed throughout the prestellar phase (the first three rows of 
Fig.~\ref{phi3}); the outer layers start to move inwards at speeds in 
the range $(0.05,0.07) \, {\rm km} \, {\rm s}^{-1}$, whilst 
matter near the centre is approximately stationary. Eventually 
the core becomes unstable, and the sink starts to accrete matter, 
marking the formation of the protostar and the start of the 
Class 0 phase. At this stage the core radius is $\sim$ 0.05 pc. 
A freefall velocity field then develops, as material 
accretes onto the protostar (fourth and fifth rows of Fig.~\ref{phi3}). The accretion 
rate reaches a maximum of $\sim 1.08 \times 10^{-5} \,M_\odot \, 
{\rm yr}^{-1}\,$, and then declines monotonically (second row of 
Fig.~\ref{accret}).

\begin{figure*}
\setlength{\unitlength}{1mm}
\begin{picture}(160,225)
\includegraphics{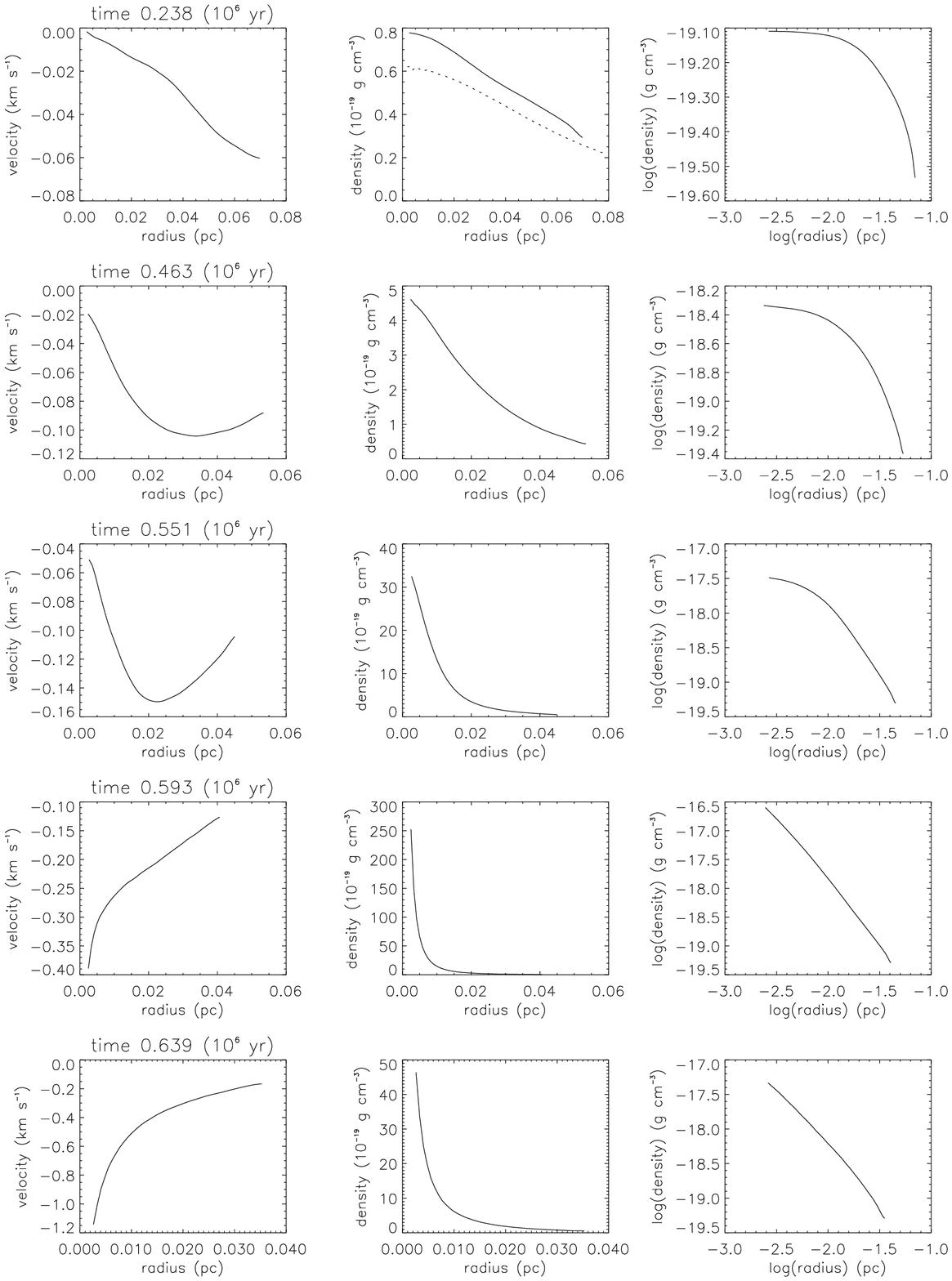}
\end{picture}
\caption{As for Fig.~\ref{phi10}, but for the case 
$\phi = 1.0$ (approximately sonic compression): 
first row, $t = 0.24 \,\mbox{\small Myrs}$; 
second row, $t = 0.46 \,\mbox{\small Myrs}$; 
third row, $t = 0.55 \,\mbox{\small Myrs}$; 
fourth row, $t = 0.59 \,\mbox{\small Myrs}$;
fifth row, $t = 0.64 \,\mbox{\small Myrs}$.}
\label{phi1}
\end{figure*}

In Figure~\ref{phi1} we show results for $\phi = 1\,$. Here the 
pressure doubles in one sound-crossing time, and so the evolution is 
more dynamic (than in the $\phi = 10$ and $\phi = 3$ cases described 
above). A small compression wave is driven into the core increasing 
the density and leaving in its wake a modest inward velocity 
field in the range $(0.10,0.15) \, {\rm km} \, {\rm s}^{-1}\,$. 
The prestellar phase ends -- and the Class 0 phase begins -- when 
this compression wave impinges on the centre. Up until this stage, 
the central density has hardly changed, since the inner parts 
have been unaware of the increased external pressure; the radius 
of the core has decreased to $\sim$ 0.045 pc. During the Class 0 
phase, a freefall velocity field develops 
around the central protostar, but the outer parts of the envelope 
are still moving inwards at approximately uniform sonic speed $(v \sim 
0.12 \, {\rm km} {\rm s}^{-1})$. The accretion 
rate is significantly higher than for the more quasistatic cases (larger 
$\phi$), reaching a maximum of $\sim 1.50 \times 10^{-5} M_\odot 
\, {\rm yr}^{-1}\,$, and then decreasing into the Class I 
phase (third row of Fig.~\ref{accret}).

\begin{figure*}
\setlength{\unitlength}{1mm}
\begin{picture}(160,225)
\includegraphics{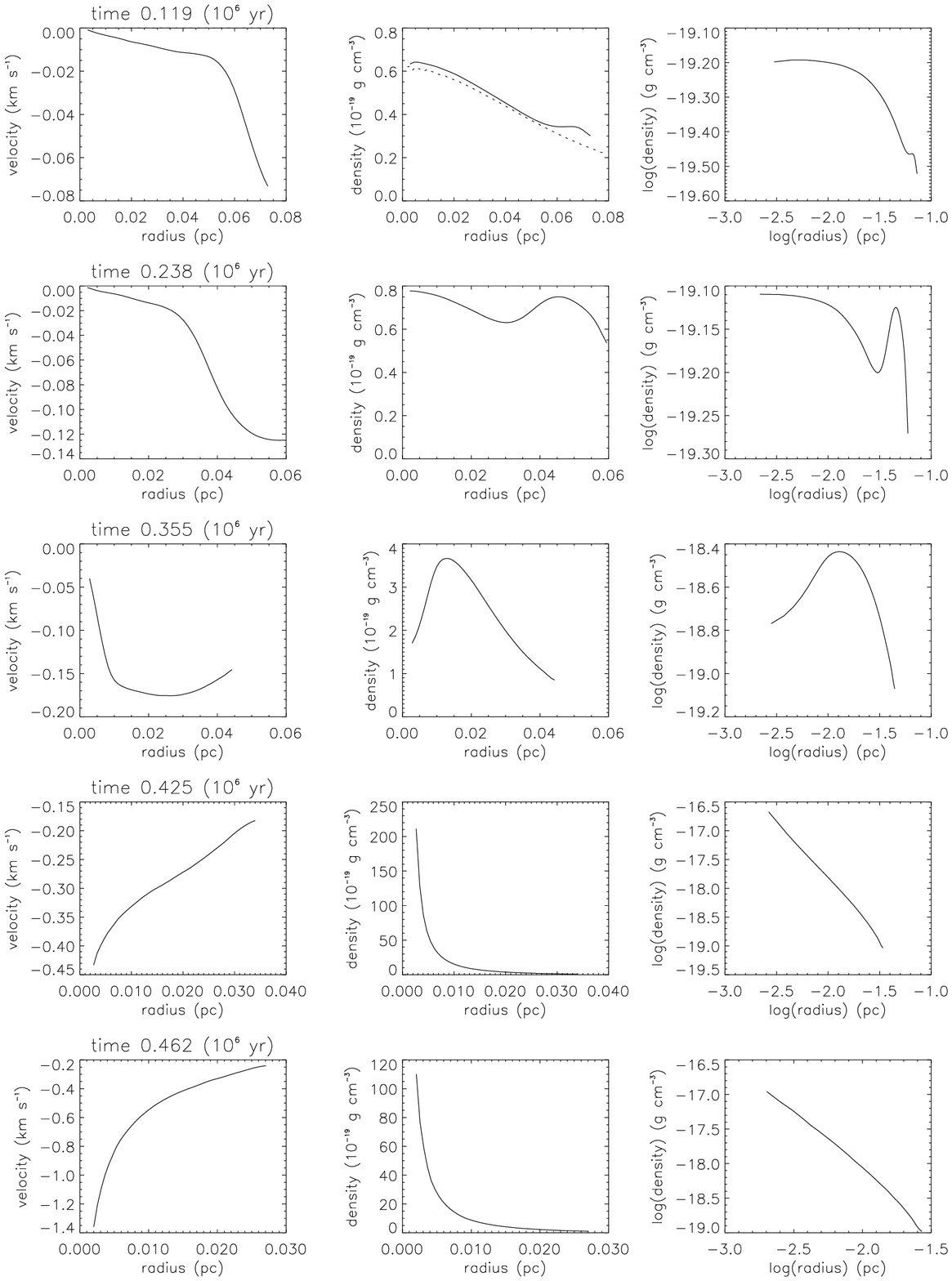}
\end{picture}
\caption{As for Fig.~\ref{phi10}, but for the case 
$\phi = 0.3$ (mildly supersonic compression): 
first row, $t = 0.12 \,\mbox{\small Myrs}$; 
second row, $t = 0.24 \,\mbox{\small Myrs}$; 
third row, $t = 0.36 \,\mbox{\small Myrs}$; 
fourth row, $t = 0.43 \,\mbox{\small Myrs}$;
fifth row, $t = 0.46 \,\mbox{\small Myrs}$.}
\label{phi03}
\end{figure*}

Figure~\ref{phi03} shows detailed results for the case $\phi = 0.3\,$, 
i.e. quite rapid compression. In this case a strong compression 
wave is driven into the core, leaving a marginally {\it sub}sonic 
velocity field 
in its wake, i.e. $v$ in the range $(0.12,0.14) \, {\small\rm km} 
{\small\rm s}^{-1})$. This leads to quite rapid accretion during 
the Class 0 phase, reaching a maximum of $\sim 1.65 \times 10^{-5} 
M_\odot \, {\rm yr}^{-1}\,$, and then decreasing into the 
Class 0 phase (fouth and fifth rows of Fig.~\ref{accret}).

\begin{figure*}
\setlength{\unitlength}{1mm}
\begin{picture}(160,225)
\includegraphics{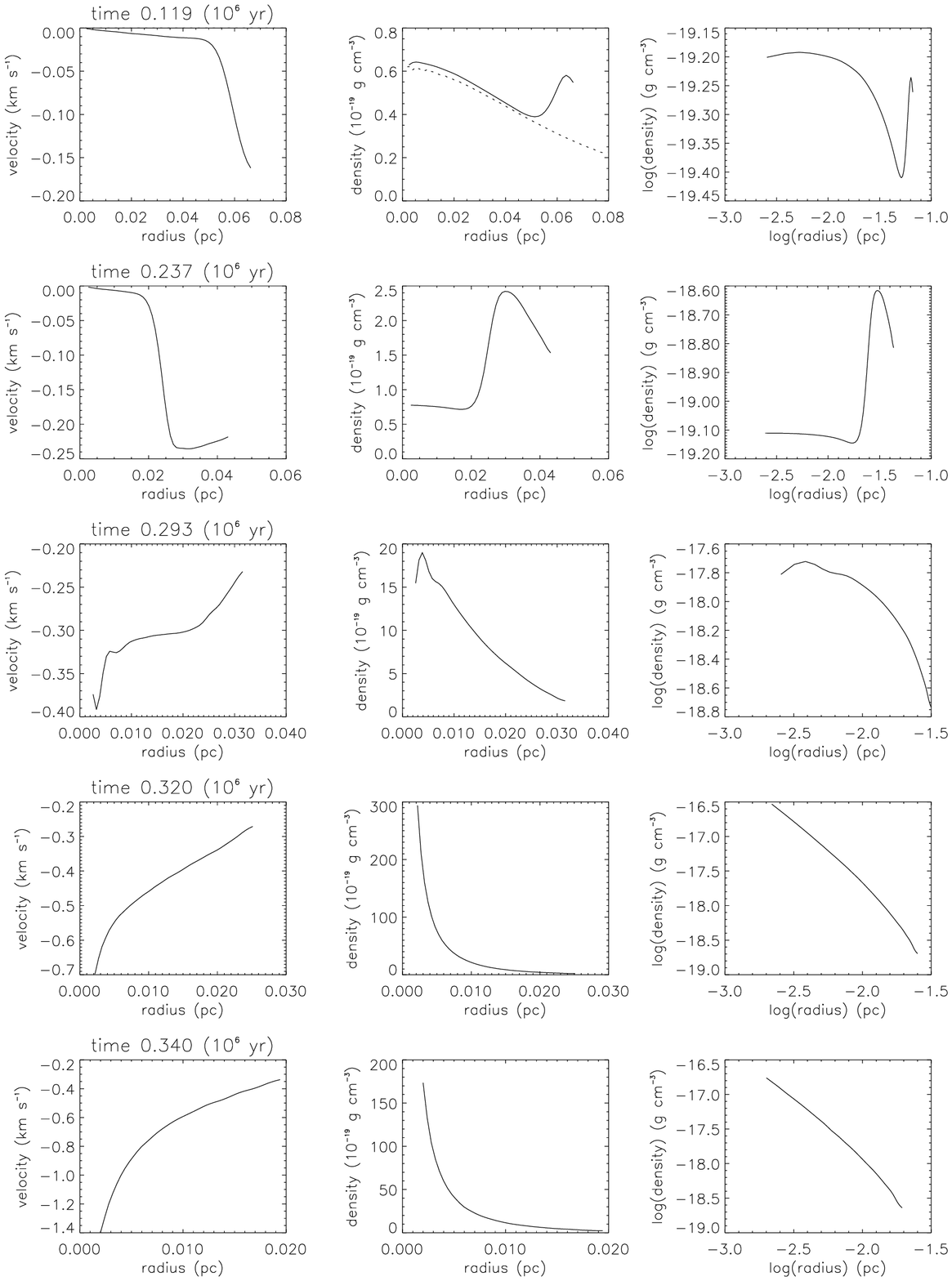}
\end{picture}
\caption{As for Fig.~\ref{phi10}, but for the case $\phi = 0.1$: 
first row, $t = 0.12 \,\mbox{\small Myrs}$; 
second row, $t = 0.24 \,\mbox{\small Myrs}$; 
third row, $t = 0.29 \,\mbox{\small Myrs}$; 
fourth row, $t = 0.32 \,\mbox{\small Myrs}$;
fifth row, $t = 0.34 \,\mbox{\small Myrs}$.}
\label{phi01}
\end{figure*}

Figure~\ref{phi01} shows detailed results for the case $\phi = 0.1\,$, 
i.e. very rapid compression. In this case a very strong compression 
wave is driven into the core, leaving a mildly {\it super}sonic velocity field 
in its wake, i.e. $v$ in the range $(0.25,0.30) \, {\small\rm km} 
{\small\rm s}^{-1})$. At $t = 0.24 \, {\rm Myr}\,$, the outer 
layers have been swept up into a {\it tsunami}. The density at the head 
of the {\it tsunami} is $\sim 3$ times higher than at the centre. 
This leads to very rapid accretion during 
the Class 0 phase, reaching a maximum of $\sim 2.60 \times 10^{-5} 
M_\odot \, {\rm yr}^{-1}\,$, and then decreasing into the 
Class I phase (fifth row of Fig.~\ref{accret}). In the Class 0 phase, a freefall 
velocity field develops in the centre, whilst the outer layers continue 
to cruise inwards at $v \sim 0.30 \, {\small\rm km} {\small\rm s}^{-1}$.

\begin{figure*}
\setlength{\unitlength}{1mm}
\begin{picture}(160,225)
\includegraphics{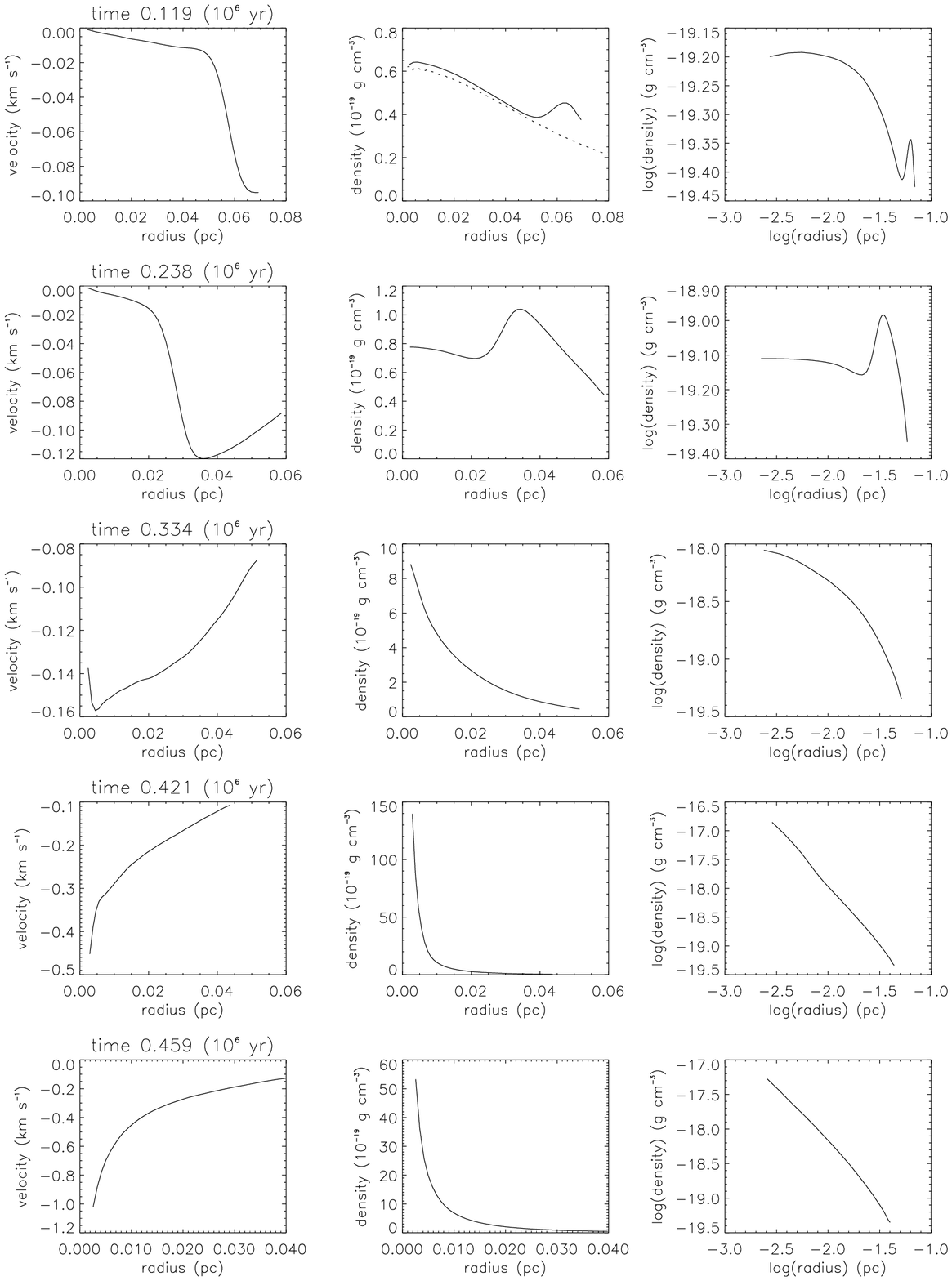}
\end{picture}
\caption{As for Fig.~\ref{phi10}, but for the case $\phi = 0.1\,$(finite), 
where instead of letting $P\xt$ increase indefinitely, the increase is halted 
once $P\xt$ has doubled: 
first row, $t = 0.12 \,\mbox{\small Myrs}$; 
second row, $t = 0.24 \,\mbox{\small Myrs}$; 
third row, $t = 0.33 \,\mbox{\small Myrs}$; 
fourth row, $t = 0.42 \,\mbox{\small Myrs}$;
fifth row, $t = 0.46 \,\mbox{\small Myrs}$.}
\label{phi01lim}
\end{figure*}

If the external pressure only increases for a finite time, and then 
stays constant at a new higher value, the results are somewhat changed, 
particularly for the cases of rapid compression (small $\phi$). This is 
illustrated on Fig.~\ref{phi01lim}, where we show detailed results for the $\,\phi=0.1\,$ 
case when the pressure increase is halted as soon as the pressure has doubled. 
We refer to this case as ``$\phi=0.1\,$(finite)'', to distinguish it from the 
standard case ``$\phi=0.1\,$(indefinite)''. In the case ``$\phi=0.1\,$(finite)'', 
the compression wave is much weaker and slower than for the case 
``$\phi=0.1\,$(indefinite)'', and so the inward velocity field it engenders is 
also weaker. In fact, the evolution for case ``$\phi=0.1\,$(finite)'' is more 
like that for the case ``$\phi=0.3\,$(indefinite)''; 
in particular, the compression wave takes a comparable time to converge onto 
the centre. For this reason, the mass and accretion rate for the 
case ``$\phi=0.1\,$(finite)'' are displayed with dashed lines on the same 
plot as the results for the case ``$\phi=0.3\,$(indefinite)'' (i.e. the 
fourth row of Fig.~\ref{accret}). For slower 
compression rates, $\phi \ga 1\,$, halting the pressure increase when the 
pressure has doubled has little effect, because instability is triggered 
before the pressure doubles, and the collapse dynamics of the core are 
therefore already well established.

To test for convergence, we have repeated the case ``$\phi=1\,$(indefinite)'' 
with only ${\cal N}_{\rm int}=20,000$ particles in the core. The 
time-dependence of the protostellar mass and accretion rate are shown on the 
third row of Fig.~\ref{accret} with dotted lines. We see that there is almost no change, 
except that with fewer particles the compression wave takes a little longer to 
converge on the centre, about 3\% longer. We do not know the reason for 
this, but in any case it is a very small effect. Apart from this time 
delay, the detailed density and velocity profiles are essentially unchanged.

\begin{table}
\caption{The durations of the prestellar and Class 0 phases ($\Delta t_{prestellar} \;\, {\rm and} \;\, \Delta t_{\rm Class 0}$), and the mean magnitude 
of the inward velocity field set up during the prestellar phase phases, 
$\bar{v}_{\rm cruise}$, for different values of $\phi\,$.}
\begin{center}\begin{tabular}{rccc}
$\phi$ & $\Delta t_{prestellar}$ & $\Delta t_{\rm Class 0}$ & 
$\bar{v}_{\rm cruise}$ \\
 & (Myrs) & (Myrs) & (km s$^{-1}$) \\
10.0 & 1.2 & 0.11 & 0.025 \\
3.0 & 0.8 & 0.05 & 0.06 \\
1.0 & 0.6 & 0.04 & 0.12 \\
0.3 & 0.4 & 0.03 & 0.17 \\
0.1 & 0.3 & 0.02 & 0.30 \\
\end{tabular}\end{center}
\end{table}

Figure~\ref{accret} shows the accretion rate onto the sink, as a function 
of time, for the various values of $\phi$. The star indicates 
where the Class 0 phase gives way to the Class I phase. 
Table 1 shows the durations of the prestellar and 
Class 0 phases, for representative values of $\phi\,$. 
Since the sink particle is present from the outset, the moment of 
protostar formation is not precisely defined, and therefore these 
durations are not precisely defined. They simply illustrate the 
fact that the prestellar and Class 0 phases are 
accelerated by rapid compression. In all cases, $\Delta t_{\rm Class 0} 
/ \Delta t_{prestellar} \sim 0.05\,{\rm to}\,0.10$.

We also tabulate values of $\,\bar{v}_{\rm cruise}\,$, 
which is the mean inward velocity established in the outer 
parts of the core as a result of the passage of the inward propagating 
compression wave. These values are obtained from the detailed 
velocity profiles by inspection, and are therefore only 
indicative of the general trend. The inward velocities observed 
in the outer envelopes of isolated prestellar cores in nearby star formation regions range from $\sim 0.04\,{\rm km}\,
{\rm s}^{-1}$ to $\sim 0.10\,{\rm km}\,{\rm s}^{-1}$ (Lee, Myers 
\& Tafalla 1999). Our models generate comparable velocities for 
$\phi$ in the range $(1,10)$, i.e. for situations in which the 
external pressure increases on a time-scale comparable with or 
greater than the sound-crossing time.

We speculate that collapses with lower values of $\phi$, i.e. 
more dynamical collapses, arise in more violently triggered star 
formation regions such as the $\rho$ Ophiuchi and Perseus protoclusters
(cf. Motte \& Andr\'e 2001). This would explain their having higher 
absolute densities and smaller extents at the onset of the Class 0 phase.

\section{Discussion}

The obvious inference to be made from these simulations is the 
unsurprising one that, the more rapid and the more prolonged 
the increase in external 
pressure, the more rapid is the evolution, the stronger is the 
compression wave, the higher is the peak accretion rate, and the 
shorter are the prestellar and Class 0 phases. However, other 
important inferences are to be found in the detail. In 
particular, simulations with {\bf $\phi \ga 1\,$} (i.e. pressure 
increasing on a sound-crossing time or slower) match essentially 
all the observational constraints described in Section 2.

First, the density profiles in the prestellar phase are relatively 
flat in the inner regions, $\eta \equiv - d \ell n [\rho] / d \ell n[r] 
\la 1 \,$, and steepen towards $\eta \sim 2\,$ in the outer 
parts. The effect of the compression wave will be hard to 
detect in projection. It will simply produce an apparent 
extension of the relatively flat inner part of a prestellar core. 
By the Class 0 and Class I phases, the density profiles should steepen, 
viz. to $\eta \sim 3/2\,$ in the inner regions, and to $\eta \sim 2-3\,$ 
in the outermost regions (see log-log plots of the density at late 
times in Figs.~\ref{phi10} to~\ref{phi01}.

Second, the velocity fields are very reminiscent of that inferred 
for L1544 by Tafalla et al. (1998) and Williams et al. (1999). In 
L1544 the velocity field in the outer parts of the core appears to 
be approximately uniform, with magnitude $\sim 0.08\,{\rm km}\,
{\rm s}^{-1}\,$. Other cores showing similar infall signatures give 
inward velocities in the range $\sim 0.04\,{\rm km}\,
{\rm s}^{-1}$ to $\sim 0.10\,{\rm km}\,{\rm s}^{-1}$ (Lee, Myers 
\& Tafalla 1999). If these velocities are due to an inward-driven 
compression wave of the type we have simulated, it suggests that 
the external pressure increases on a time-scale comparable with, but 
somewhat greater than, a sound-crossing time.

The generation of an approximately uniform inward radial velocity 
field appears to be a generic feature of compression waves. All of 
the two-dimensional infinity of similarity solutions found by 
Whitworth \& Summers (1985) involve compression waves which are driven from 
the outside inwards, and which leave behind them a uniform 
inward velocity field ($v_{\rm rad} \simeq \; 
{\rm constant}$) and a $\rho \propto r^{-2}$ density field. 
The convergence of this compression wave on the centre at $t = 0$ 
signals the formation of a central stellar object. This object 
then grows at a constant accretion rate. At the same time 
an expansion wave is reflected outwards and leaves in its wake 
an approximately freefall velocity field, $v_{\rm rad} \propto 
r^{-1/2}\,$, and $\rho \propto r^{-3/2}\,$. The Shu (1977) solution is 
simply the limiting case of a zero-amplitude compression wave 
leaving a zero-amplitude inward velocity field in its wake; hence, 
uniquely, it starts from a static singular isothermal sphere. The 
Larson-Penston solution involves a fairly strong compression wave 
propagating into a uniform-density, homologously contracting cloud. 
If radial velocity fields like that inferred for L1544 turn out 
to be common, one possible explanation is that core collapse and protostar
formation generally follow from external compression. 

Third, the accretion rate is very high immediately after the 
formation of the central protostar, i.e. in the Class 0 phase, 
but later it decreases. This conforms to the pattern of 
accretion inferred from the decline in outflow power between the 
Class 0 and the Class I phase (Bontemps et al. 1996, Henriksen et 
al. 1997), and is consistent with the relative numbers of Class 0 and 
Class I sources.

A rapid increase of external pressure, of the sort we have 
invoked in these simulations, could arise in a number of 
situations. In the first instance, we might presuppose the 
existence of an 
approximately stable gravitationally bound core. If this 
core were then overrun by a strong shock which propagated 
supersonically through the medium surrounding the core, the 
external pressure acting on the core boundary would be increased. 
Alternatively, if the medium surrounding the core were suddenly 
irradiated more intensely, because a luminous star switched on 
nearby, or came out from behind a dust cloud, the surrounding medium would 
be heated -- and possibly also ionized -- giving a rapid increase 
in the external pressure. 
A contraction of the molecular cloud in which the core is embedded, 
as in the scenario of Palla \& Stahler (2000), 
would also lead to an increase (presumably slow) in the  
external pressure bounding the dense core. 
A similar type of collapse would ensue if the radiation 
heating the core were suddenly shut off (but not that heating 
the external medium). The core would become thermally unstable, 
and the external medium would drive in a compression wave. 
It is also possible that the same convergent flows which 
create cores in the first place, may sometimes then continue to 
compress these cores until they collapse, as in the scheme recently 
described by  Hartmann, Ballesteros-Paredes \& Bergin (2001).

However, it is very unlikely that the increase in the external pressure 
would actually be isotropic -- as we have assumed, for simplicity, 
in the simulations presented here. Therefore we should not expect 
compression waves to converge on the centre simultaneously from 
all directions. Instead, the compression waves arriving from 
different directions are likely to be out of phase, so that they 
will interfere in complex ways. This could have the effect of 
creating multiple protostellar seeds in the interior of a core, in 
accordance with the observation that most stars are born in 
binary or higher multiple systems (Mathieu 1994).

\section{Conclusions}

We conclude that the model of protostellar collapse developed 
here merits further investigation. It appears to reproduce the 
gross features of the density and velocity fields observed in 
prestellar cores and protostars, and the relative ages of 
the Class 0 and Class I phases. We now need to explore the 
consequences of introducing an energy equation, the 
effect of initial rotation, and the influence of 
perturbations as a means of inducing the formation of binaries 
and higher multiples.

\section*{Acknowledgements}

PH, APW{, and PhA} gratefully acknowledge the support of an 
European Commission 
Research Training Network under the Fifth Framework Programme (No. 
HPRN-CT2000-00155), and useful discussions with Derek Ward-Thompson, 
Arnaud Belloche and Pierre Lesaffre. PPG gratefully acknowledges the 
support of a PPARC postgraduate studentship.

\bsp

\label{lastpage}

\end{document}